\documentclass[12pt, preprint, amsmath,amssymbre]{revtex4}
\usepackage{amsmath}
\usepackage{amsfonts}
\usepackage{amssymb}
\usepackage[colorlinks,linkcolor=blue,citecolor=red]{hyperref}
\usepackage{graphicx}
\usepackage{graphics}
\usepackage{hyperref}
\usepackage{subfigure}
\usepackage{graphicx}
\usepackage{dcolumn}
\usepackage{bm}
\usepackage{amsmath,amsthm,amsfonts}
\usepackage{graphicx}
\usepackage{physics}
\usepackage{hyperref}
\usepackage{lscape} 

\usepackage{mathtools}
\usepackage{appendix}
\usepackage{comment}
\graphicspath{{figs/}}

\newcommand{\nc}{\newcommand}
\nc{\ba}{\begin{eqnarray}}
\nc{\ea}{\end{eqnarray}}

\nc{\rc}{\textcolor[rgb]{1.00,0.00,0.00}}
\nc{\bc}{\textcolor[rgb]{0.00,0.07,1.00}}
\nc{\bfk}{\bf{k} }
\nc{\calR}{{\cal{R}}}
\nc{\calP}{{\cal{P}}}
\nc{\bfq}{{\bf{q}}}
\nc{\bfp}{{\bf{p}}}
\nc{\Ha}{{\bf H}_3  }
\nc{\Hb}{{\bf H}_4  }

\nc{\cN}{ {\cal{N}} }

\begin{document}

\title{Stochastic Inflation with Interacting Noises}


\author{Amin Nassiri-Rad$^{a}$}
\email{amin.nassiriraad@kntu.ac.ir}

\author{Haidar Sheikhahmadi$^{b}$}
 \email{h.sh.ahmadi@gmail.com}

\author{Hassan Firouzjahi$^{b}$}
 \email{firouz@ipm.ir}


  \affiliation{$^{a}$ Department of physics, K.N Toosi University of Technology, P.O. Box 15875-4416, Tehran, Iran}
  \affiliation{$^{b}$ School of Astronomy, Institute for Research in Fundamental Sciences (IPM), P. O. Box 19395-5746, Tehran, Iran\\
  \\
  \\}
\begin{abstract}
Stochastic $\delta N$ formalism is a powerful tool to calculate the cosmological correlators non-perturbatively. However, it requires the initial data for the amplitude of the noise on the initial flat hypersurface which for a free theory
during inflation is  fixed to be $\frac{H}{2 \pi}$. In this work, we study the setups where the underlying theory involves interactions and the stochastic noises inherit these interactions.
We extend the stochastic $\delta N$ formalism to these setups  and  rewrite the corresponding Langevin and Fokker-Planck equations in which the QFT corrections in the amplitude of the noises are taken into account. 
As an example, in the three-phase SR-USR-SR  setup which is  employed for PBHs formation, the modification in the amplitude of noise is calculated from the one-loop corrections in power spectrum via in-in formalism. We show that in these setups the amplitude of the 
stochastic noise is modified to  $\frac{H}{2 \pi} \Big(1+ \frac{ \Delta {\cal P}_{\cal R} }{ {\cal P}^{(0)}_{ {\cal R} } }\Big)^{\frac{1}{2}}$ in which  $ \frac{\Delta {\cal P}_{\cal R} }{ {\cal P}^{(0)}_{{\cal R} } }$ is the fractional one-loop correction in power spectrum. 
\end{abstract}

\maketitle

\newpage

\section{Introduction}\label{introduction}
Cosmological inflation is the widely accepted paradigm to address the
horizon and flatness problems associated with big bang cosmology
on large scales. This process occurs at high energies in the early universe. 
Furthermore, inflation can address  the origin of the seeds of the large-scale structure  through quantum vacuum fluctuations  \cite{Weinberg:2008zzc, Baumann:2009ds}.

The quantum fluctuations are characterized by curvature perturbations with  a small amplitude $\mathcal{O}(10^{-5})$ on CMB scales  \cite{Planck:2018vyg,Planck:2018jri} but may attain larger values on smaller scales. Upon exiting the Hubble radius, these perturbations become classical and can be treated stochastically. The stochastic formalism thus offers a robust framework for modelling their evolution \cite{Vilenkin:1983xp, Starobinsky:1986fx, Rey:1986zk, Nakao:1988yi, Sasaki:1987gy, Nambu:1987ef, Nambu:1988je, Starobinsky:1994bd,Kunze:2006tu,Prokopec:2007ak,Prokopec:2008gw,Tsamis:2005hd, Enqvist:2008kt,Finelli:2008zg,Finelli:2010sh, Garbrecht:2013coa, Garbrecht:2014dca, Burgess:2014eoa,Burgess:2015ajz,Cohen:2021fzf, Boyanovsky:2015tba,Boyanovsky:2015jen, Cruces:2021iwq, Cruces:2024pni, Ahmadi:2022lsm, Noorbala:2019kdd, Noorbala:2024fim, Cable:2022uwd, Cable:2023gdz, Mizuguchi:2024kbl, Murata:2025onc, Mishra:2023lhe, Jackson:2024aoo}. In this formalism, the quantum fluctuations of scalar fields (such as the inflaton) in Fourier space are decomposed into long- and short-wavelength modes. The long-wavelength modes encode the evolution of the background field, while the short-wavelength modes affect the long-wavelength modes as classical noises upon horizon exit. For Gaussian fluctuations, the amplitude of these quantum jumps is given by $H/2\pi$, where $H$ denotes the Hubble expansion rate during inflation. The stochastic $\delta N$ formalism offers a systematic approach to compute the correlation functions of these quantum fluctuations \cite{Fujita:2013cna, Fujita:2014tja, Vennin:2015hra,Vennin:2016wnk, Assadullahi:2016gkk,Grain:2017dqa,Noorbala22,Firouzjahi:2018vet, Firouzjahi:2020jrj}. This method is rooted in the $\delta N$ formalism, a robust framework for computing curvature perturbations non-perturbatively (non-linearly)  \cite{Sasaki:1995aw, Sasaki:1998ug,  Wands:2000dp, Lyth:2004gb, Lyth:2005fi, Cruces:2025typ}. The $\delta N$ formalism arises naturally from the separate universe approach where superhorizon modes evolve independently within causally disconnected patches and influence only their local neighbourhoods.

A key advantage of the stochastic formalism is its non-perturbative nature which enables the description of large field fluctuations in non-linear regime \cite{Enqvist:2008kt,Asadi:2023flu,Fujita:2014tja}. In such cases, the energy density of fluctuations can grow sufficiently large to produce primordial black holes (PBHs). One scenario where curvature perturbations may be amplified large enough to form PBHs is the ultra-slow-roll (USR) inflation, characterized by an extremely flat potential \cite{Kinney:2005vj,Namjoo:2012aa, Chen:2013aj, Chen:2013eea, Martin:2012pe, Morse:2018kda, Motohashi:2014ppa, Dimopoulos:2017ged, Pi:2022ysn}. The abundance of PBHs across different mass ranges is determined by the tail of the probability distribution function (PDF) for curvature perturbations during their enhancement phase \cite{Pattison:2018bct,Biagetti:2018pjj,Garcia-Bellido:2017mdw,Ivanov:1994pa, Pattison:2017mbe,Germani:2017bcs, Ezquiaga:2018gbw,Ezquiaga:2019ftu,Pattison:2021oen,Hooshangi:2023kss}. As a non-perturbative approach, the stochastic formalism provides a powerful tool for determining the PDF of field fluctuations. 

The standard picture employing USR mechanism for PBHs formation is the setup with three phases of inflation, SR-USR-SR. The first stage is a SR phase during which the observable CMB perturbations are generated. The second stage is a USR phase during which the perturbations are amplified to generate PBHs at the desired scales.
The USR phase lasts for a few e-folds followed by the final SR phase where the system reaches its attractor phase and inflation ends via reheating.
An important question is  whether or not the long CMB scales perturbations are affected by small scales USR modes as for example advocated in \cite{Kristiano:2022maq}. { To study this question from the viewpoint of stochastic inflation, a key issue is that  the statistics of scalar field fluctuations are not strictly Gaussian. These quantic loop corrections in scalar spectrum modify the noise, causing it to be interacting and consequently  a modified Gaussian one \footnote{ {Here by modified Gaussian, we mean the amplitude of the noises is modified compared to the free theory.} }}. One main goal of this work is to formulate the stochastic inflation  where the noises are not free and involve interaction in the matter sector. To properly account for these effects, we propose modifying the noise amplitude in the stochastic formalism using the standard QFT  methods. As we will show, the modification of the noise term directly affects the PDF with non-trivial applications.

\section{Review of Stochastic Inflation}
\label{sec:stochastic_inflation}

In this section, we  briefly review the formalism of  stochastic inflation which
serves as the  basis for our subsequent analysis.

We consider a single-field inflationary model described by a scalar potential $V(\phi)$. The dynamics of the scalar field $\phi$ is governed by the Klein-Gordon equation in an expanding Universe,
\begin{equation}
\label{eq:klein_gordon}
    \left(\frac{\partial^2}{\partial t^2} + 3H\frac{\partial}{\partial t} - \frac{\nabla^2}{a^2}\right)\phi(\mathbf{x}, t) + \frac{\partial V}{\partial \phi}(\mathbf{x}, t) = 0,
\end{equation}
in which $a(t)$ is the FLRW scale factor with $t$ being the cosmic time
and $H= \dot a/a$ is the Hubble expansion rate during inflation.

In this formulation, we have neglected the metric perturbations and gravitational back-reactions. This simplification is justified in the stochastic $\delta N$ formalism where initial quantum fluctuations are evaluated on spatially
flat hypersurfaces, and upon choosing an appropriate gauge, the spatial metric perturbations can be set to zero. Additionally, within the Arnowitt-Deser-Misner (ADM) decomposition, the lapse and shift functions appear only at higher orders in the gradient expansion. Their effects are therefore subleading  and can be omitted safely~\cite{Lyth:2004gb}. This allows us to focus exclusively on the scalar field dynamics to leading order in gradient expansion.

To analyze the behaviour of fluctuations, we decompose the scalar field $\phi$ and its conjugate momentum $\Pi$ into long-wavelength and short-wavelength components. This decomposition is physically motivated by the fact that the sub-horizon modes oscillate rapidly and typically average out over time, whereas super-horizon modes evolve slowly and contribute significantly to the large-scale structure of the Universe. The long-short
decompositions for the scalar field and its conjugate momenta  are expressed as follows,
\begin{equation}
\label{eq:phi_decomposition}
    \phi(\mathbf{x}, t) = \phi_l(\mathbf{x}, t) + \sqrt{\hbar} \, \phi_s(\mathbf{x}, t),
\end{equation}
and
\begin{equation}
\label{eq:Pi_decomposition}
    \Pi(\mathbf{x}, t) = \Pi_l(\mathbf{x}, t) + \sqrt{\hbar} \, \Pi_s(\mathbf{x}, t),
\end{equation}
where the subscripts $l$ and $s$ denote the long-wavelength and short-wavelength components respectively. The quantum nature of the short modes is made explicit in the above decompositions by the factor of $\sqrt{\hbar}$. These short modes are further expanded in Fourier space as follows \cite{Nakao:1988yi,Sasaki:1987gy},
\begin{equation}
\label{eq:phi_s_fourier}
    \phi_s(\mathbf{x}, t) = \int \frac{d^3k}{(2\pi)^3} \, \theta(k - \varepsilon aH) \, \phi_{\mathbf{k}}(t) e^{i\mathbf{k} \cdot \mathbf{x}},
\end{equation}
\begin{equation}
\label{eq:Pi_s_fourier}
    \Pi_s(\mathbf{x}, t) = \int \frac{d^3k}{(2\pi)^3} \, \theta(k - \varepsilon aH) \, \dot{\phi}_{\mathbf{k}}(t) e^{i\mathbf{k} \cdot \mathbf{x}}.
\end{equation}
Here, $\theta$ is the Heaviside step function and
$\varepsilon$ is a small dimensionless parameter ($\varepsilon \ll 1$) used to separate the long and short scales in a controlled way which should not be confused with the first SR parameter $\epsilon_H$.   The Fourier coefficients $\phi_{\mathbf{k}}$ are given in terms of creation and annihilation operators as $\phi_{\mathbf{k}} = a_{\mathbf{k}} \varphi_k + a^\dagger_{-\mathbf{k}} \varphi_{-k}^*$, where $\varphi_k$ denotes the positive-frequency mode functions.

Expanding the Klein-Gordon equation~\eqref{eq:klein_gordon} around $\phi_l$ and $\Pi_l$ to first order in $\sqrt{\hbar}$, we obtain the equations of motion for the long-wavelength components as follows
\cite{Nakao:1988yi,Sasaki:1987gy},
\begin{align}
\label{eq:phi_l_dot}
    \dot{\phi}_l &= \Pi_l + \sqrt{\hbar} \,  \xi_\phi, \\
\label{eq:Pi_l_dot}
    \dot{\Pi}_l &= -3H \Pi_l + \frac{1}{a^2} \nabla^2 \phi_l - V'(\phi_l) + \sqrt{\hbar} \,  \xi_\Pi,
\end{align}
where $\xi_\phi$ and $\xi_\Pi$ are stochastic noise terms induced by the short-wavelength modes,
\begin{equation}
\label{eq:sigma}
    \xi_\phi(\mathbf{x}, t) \equiv \varepsilon aH^2 \int \frac{d^3k}{(2\pi)^3} \, \delta(k - \varepsilon aH) \, \phi_{\mathbf{k}}(t) e^{i\mathbf{k} \cdot \mathbf{x}},
\end{equation}
and,
\begin{equation}
\label{eq:tau}
   \xi_\Pi(\mathbf{x}, t) \equiv \varepsilon aH^2 \int \frac{d^3k}{(2\pi)^3} \, \delta(k - \varepsilon aH) \, \dot{\phi}_{\mathbf{k}}(t) e^{i\mathbf{k} \cdot \mathbf{x}}.
\end{equation}
These noise terms act as sources for the dynamics of the long-wavelength fields, effectively capturing the stochastic influence of quantum fluctuations as they cross the horizon.

Assuming free Gaussian statistics for the field fluctuations and adopting the Bunch-Davies vacuum $|0\rangle$, the noise correlators are given by~\cite{Nakao:1988yi, Sasaki:1987gy}, 
\begin{align}
\label{eq:sigma_correlation}
    \langle 0 | \xi_\phi(\mathbf{x}_1) \xi_\phi(\mathbf{x}_2) | 0 \rangle &\approx \varepsilon^{\frac{2M^2}{3H^2}} \frac{H^3}{4\pi^2} j_0(\varepsilon aH |\mathbf{x}_1 - \mathbf{x}_2|) \delta(t_1 - t_2), \\
\label{eq:tau_correlation}
    \langle 0 | \xi_\Pi(\mathbf{x}_1) \xi_\Pi(\mathbf{x}_2) | 0 \rangle &\approx \varepsilon^{\frac{2M^2}{3H^2}} \left(\frac{M^2}{3H^2} + \varepsilon^2\right)^2 \frac{H^5}{4\pi^2} j_0(\varepsilon aH |\mathbf{x}_1 - \mathbf{x}_2|) \delta(t_1 - t_2), \\
\label{eq:sigma_tau_correlation}
    \langle 0 | \xi_\phi(\mathbf{x}_1) \xi_\Pi(\mathbf{x}_2) + \xi_\Pi(\mathbf{x}_2) \xi_\phi(\mathbf{x}_1) | 0 \rangle &\approx -2 \varepsilon^{\frac{2M^2}{3H^2}} \left(\frac{M^2}{3H^2} + \varepsilon^2\right) \frac{H^4}{4\pi^2} j_0(\varepsilon aH |\mathbf{x}_1 - \mathbf{x}_2|) \delta(t_1 - t_2),
\end{align}
where $M^2$ denotes the effective mass of the long-wavelength field, and $j_0$ is the zeroth-order spherical Bessel function.

Finally, the commutation relations for the noise terms are,
\begin{align}
\label{eq:sigma_commutation}
    [\xi_\phi(\mathbf{x}_1), \xi_\phi(\mathbf{x}_2)] &= [\xi_\Pi(\mathbf{x}_1), \xi_\Pi(\mathbf{x}_2)] = 0, \\
\label{eq:sigma_tau_commutation}
    [\xi_\phi(\mathbf{x}_1), \xi_\Pi(\mathbf{x}_2)] &= i \varepsilon^3 \frac{H^4}{4\pi^2} j_0(\varepsilon aH |\mathbf{x}_1 - \mathbf{x}_2|) \delta(t_1 - t_2).
\end{align}
As seen above, the commutators vanish in the limit $\varepsilon \to 0$, illustrating that the noise terms become effectively classical. This marks the transition from quantum fluctuations to classical stochastic behaviour, an important step  in the formulation of the stochastic inflation.

\subsection{USR Setup}

The formalism presented above is  general and does not rely on a specific form of the inflaton potential with the assumption that the field perturbations are free and Gaussian.   As a simple example, consider the case of purely USR inflation, characterized by a constant potential $V(\phi) = V_0$. In this regime, the effective mass of the inflaton field vanishes, i.e. $M^2 = 0$ and the first SR parameter $\epsilon_H$ decays exponentially as $a^{-6}$. Furthermore, 
from Eqs.~\eqref{eq:sigma_correlation}--\eqref{eq:sigma_tau_correlation}, it follows that in the limit $M^2 \to 0$, the $\varepsilon$-dependence of the noise term $\xi_\phi$ disappears, while $\xi_\Pi = \mathcal{O}(\varepsilon^2)$. Consequently, for sufficiently small $\varepsilon$, $\xi_\Pi$ becomes negligible and can be dropped. The leading noise contribution is thus solely due to $\xi_\phi$, and the corresponding correlation function simplifies to,
\begin{equation}
\label{eq:sigma_correlation_USR}
    \langle 0 | \xi_\phi(\mathbf{x}_1) \xi_\phi(\mathbf{x}_2) | 0 \rangle \approx \frac{H^3}{4\pi^2} \delta(t_1 - t_2) = \frac{H^4}{4\pi^2} \delta(N_1 - N_2),
\end{equation}
where we have changed the clock from cosmic time $t$ to the number of $e$-folds $dN \equiv H dt$. In this limit, $\xi_\phi$ becomes spatially homogeneous and {time-independent}, a direct consequence of the flatness of the potential and the ensuing suppression of the gradient terms. Furthermore, the amplitude of the noise is set to $\frac{H}{2 \pi}$, a fundamental property of stochastic inflation in the absence of interaction.

In the super-horizon limit ($k \ll aH$), and dropping the subscript $l$ for notational clarity, Eqs.~\eqref{eq:phi_l_dot} and \eqref{eq:Pi_l_dot} in this simple example of USR setup  reduce to the following Langevin equation,
\begin{align}
\label{eq:phi_dN}
    \frac{d\phi}{dN} &= {\Pi} + \frac{H}{2\pi} \xi(N), \\
\label{eq:Pi_dN}
    \frac{d\Pi}{dN} &= -3\Pi,
\end{align}
where we have defined
{ $\xi_\phi \equiv \frac{H^2}{2\pi} \xi$}, with $\xi(N)$
 representing a normalized Gaussian white noise satisfying,
\ba
\label{eq:xi_noise_mean}
    \langle \xi(N) \rangle = 0, \quad \quad 
    \langle \xi(N) \xi(N') \rangle &= \delta(N - N').
\ea

A notable feature of the USR regime is that while $\phi$ evolves stochastically due to quantum noise, the evolution of the conjugate momentum $\Pi$ remains fully deterministic. This is because $\xi_\Pi = \mathcal{O}(\varepsilon^2)$  so 
its effect can be safely ignored.

The solution of Eq.~\eqref{eq:Pi_dN} is straightforward,
\begin{equation}
\label{eq:Pi_solution}
    \Pi(N) = {\Pi}_\text{{ini}} e^{-3N},
\end{equation}
where ${\Pi}_\text{{ini}}$ is the initial momentum of the inflaton at the onset of the USR phase, which we set to $N = 0$.

Substituting Eq.~\eqref{eq:Pi_solution} into Eq.~\eqref{eq:phi_dN}, we arrive at the Langevin equation governing the coarse-grained evolution of the inflaton \cite{Firouzjahi:2018vet},
\begin{equation}
\label{eq:langevin}
    \frac{d\phi}{dN} = {\Pi}_\text{{ini}} e^{-3N} + \frac{H}{2\pi} \xi(N).
\end{equation}
Integrating this expression with initial condition $\phi(0) = \phi_\text{{ini}}$, we obtain
\begin{equation}
\label{eq:phi_solution}
    \phi(N) = \phi_\text{{ini}} + \frac{{\Pi}_\text{{ini}}}{3} \left(1 - e^{-3N}\right) + \frac{H}{2\pi} W(N),
\end{equation}
where
\begin{equation}
\label{eq:wiener_process}
    W(N) \equiv \int_0^N \xi(N') \, dN'
\end{equation}
is the Wiener process associated with the noise $\xi(N)$~\cite{Firouzjahi:2018vet}.

In deriving Eq.~(\ref{eq:phi_solution}), we have assumed that the Hubble parameter $H$ is approximately constant throughout the USR phase. This assumption is well justified, as $\epsilon_H \propto a^{-6}$ leads to a negligible time variation in $H$. Equation~ (\ref{eq:phi_solution}) encapsulates both the classical drift due to the residual kinetic energy and the effects of quantum-induced stochastic noise.

We  emphasize that here we have assumed a pure USR phase while in the next sections we consider the example of  three-phase setup   SR-USR-SR which is employed for PBHs formation.

\subsection{ {Gaussian Noise with Modified Amplitude} and Higher-Order Corrections}

 {To study the effects of higher-order corrections to the Fokker-Planck equation induced from quantum Loop corrections, here we start with the general Fokker-Planck equation.}
As discussed in~\cite{Risken1989,Cohen:2021fzf}, in such cases the evolution of the probability distribution $P(X, t)$ is governed by a generalized Fokker-Planck equation,
\begin{equation}
\label{eq:fp_general}
\frac{\partial P(X, t)}{\partial t} = \sum_{n=1}^{\infty} \Big(-\frac{\partial}{\partial X} \Big)^n \left[ D^{(n)}(X) P(X, t) \right],
\end{equation}
with the coefficients $D^{(n)}(X)$ defined via the Kramers-Moyal expansion as,
\begin{equation}
\label{eq:kramers_moyal}
D^{(n)}(X_0) = \left. \frac{1}{n!} \lim_{\Delta t \to 0} \frac{1}{\Delta t} \Big\langle \big[X(t + \Delta t) - X(t) \big]^n \Big\rangle \right|_{t = 0}.
\end{equation}
Given that $\xi_\phi$ is defined as in Eq.~\eqref{eq:sigma}, it is straightforward to verify that all higher-order contributions with $n > 2$ are suppressed. In particular, to ensure that the expansion terminates at $n=2$ \cite{Pawula1967}, it suffices (per the Pawula theorem \cite{Pawula1967}, see appendix~\ref{AppendixB} for more details) to demonstrate that,
\begin{equation}
D^{(4)}(X) = 0.
\end{equation}
To that end,  consider the four-point correlation function,
\begin{equation}
\begin{split}
    \left\langle \xi_\phi(N_1)\xi_\phi(N_2)\xi_\phi(N_3)\xi_\phi(N_4) \right\rangle &= \int \frac{d^3 k_1}{(2\pi)^3} \cdots \frac{d^3 k_4}{(2\pi)^3} \, \delta(k_1 - \varepsilon a_1 H) \cdots \delta(k_4 - \varepsilon a_4 H) \\
    &\quad \times \left\langle \delta\phi_{k_1} \delta\phi_{k_2} \delta\phi_{k_3} \delta\phi_{k_4} \right\rangle.
\end{split}
\end{equation}
Setting $N_1 = N_2 = N_3 = N_4$ and evaluating the correlator at tree level, one finds that it scales as $\delta(0)^2$, implying that the dominant term in $D^{(4)}$ is proportional to $\delta(0)^2 \, dN^4$. Recognizing that $\delta(0) \, dN^2 \propto dN$ (as can be understood from the identity $\frac{d}{dx} \theta(x) = \delta(x)$), we conclude that $D^{(4)} = 0$, and thus all higher-order coefficients $D^{(n > 2)}$ vanish \cite{Riotto:2011sf}. Hence, the Fokker-Planck equation reduces to the standard second-order form.

It is important to note that, to avoid divergences in $D^{(1)}$, the noise must satisfy $\langle \xi_\phi(N) \rangle = 0$. While this condition is naturally satisfied in the case of Gaussian noise, it does not generally hold when the noise
involves interaction (as we consider in next Section). To remedy this, we redefine the noise term via
\begin{equation}
    \tilde{\xi}_\phi(N) = \xi_\phi(N) - \langle \xi_\phi(N) \rangle,
\end{equation}
and from now on, we work with the redefined noise variable $\tilde{\xi}_\phi$, dropping the tilde for notational simplicity.

\section{Models with Interacting Noises}
\label{Setup}

 {After this review, we consider a more general case where the noise can have intrinsic  time dependence,
either via interaction from the underlying quantum field theory or because of the effects raising from linear theory.
The main question is whether the effect induced by time-dependent noise is of the same order as the quantum loop correction. Our analysis shows that these loop corrections are, in fact, distinct from any explicit time dependence. }
We begin with the following system of Langevin equations, where the noise amplitude is assumed to be time-dependent with 
$${ \xi_\phi({N}) =  \tilde{f}(N) \xi(N) \quad \quad  \text{and} \quad \quad   \xi_\Pi({N}) =  \tilde{g}(N) \xi(N)}, $$yielding to, 
\begin{equation}
\label{Langevin}
    \frac{d\phi}{dN} = \Pi(N) + \tilde{f}(N) \xi(N), \quad
    \frac{d\Pi}{dN} = -3\Pi(N) - \frac{3V'(\phi)}{V(\phi)} + \tilde{g}(N) \xi(N),
\end{equation}
where $\xi(N)$ denotes a stochastic {Gaussian} noise term while $\tilde{f}(N)$ and $\tilde{g}(N)$ refer to the amplitude of the noise at time $N$ satisfying, 
 {
\begin{eqnarray}\label{f-ftilde} 
  \tilde{f}^2(N_1)\delta(N_1-N_2) \equiv \langle   \xi_\phi({N}_1) \xi_\phi({N}_2) \rangle,\\
    \tilde{g}^2(N_1)\delta(N_1-N_2) \equiv \langle   \xi_\Pi({N}_1) \xi_\Pi({N}_2)  \rangle.
\end{eqnarray}
}
To treat the time dependence of the noise more systematically, we introduce an auxiliary field $F$, defined by
\begin{equation}
    \frac{\partial F}{\partial N} = 1.
\end{equation}
 {The auxiliary variable $F$ is introduced to absorb the explicit time dependence of the coefficients $f$ and $g$, making the stochastic process time-homogeneous and thus allowing the use of standard first-passage-time techniques. }
Using this auxiliary variable, we rewrite the Langevin equation (\ref{Langevin}) as follows, 
\begin{equation}
\label{Langevin2}
    \frac{d\phi}{dN} = \Pi(N) + \tilde{f}(F) \xi(N), \quad
    \frac{d\Pi}{dN} = -3\Pi(N) - \frac{3V'(\phi)}{V(\phi)} + \tilde{g}(F) \xi(N), \quad
    \frac{\partial F}{\partial N} = 1.
\end{equation}
Here, we have neglected slow-roll corrections for simplicity.


{\subsection{Adjoint Fokker-Planck Equation and Characteristic Method}}

The  Fokker-Planck equation plays a fundamental role in stochastic inflation by describing the time evolution of  inflaton's PDF and tracking how initial field configurations diffuse because of the inflationary potential. The PDF obtained from the Fokker-Planck equation describes the probability of finding the field in the phase space at any given time.
On the other hand, unlike direct simulations of stochastic dynamics, the adjoint method enables efficient computation of expectation values by solving backward-time
 or the adjoint Fokker-Planck equation. The adjoint Fokker-Planck equation describes the PDF of crossing the final boundary by appropriate initial and boundary conditions.

The corresponding adjoint Fokker-Planck equation, describing the evolution of the probability distribution  {  $P(\phi, \Pi, F,  \mathcal{N})$ } is written as follows,
 {
\begin{equation}
\begin{split} \label{adjoint}
    \frac{\partial P_\mathrm{fp}}{\partial \mathcal{N}} = &
    - \left( 3\Pi + \frac{3V'(\phi)}{V(\phi)} \right) \frac{\partial P_\mathrm{fp}}{\partial \Pi}
    + \Pi \frac{\partial P_\mathrm{fp}}{\partial \phi}
    + \frac{\partial P_\mathrm{fp}}{\partial F} \\
    & + \frac{\tilde{f}(F + N_{\text{cl}})^2}{2} \frac{\partial^2 P_\mathrm{fp}}{\partial \phi^2}
    + \frac{\tilde{g}(F + N_{\text{cl}})^2}{2} \frac{\partial^2 P_\mathrm{fp}}{\partial \Pi^2}
    + \tilde{\Gamma}(F + N_{\text{cl}}) \frac{\partial^2 P_\mathrm{fp}}{\partial \phi \partial \Pi},
\end{split}
\end{equation}
with the boundary condition
\begin{equation} \label{Boundary}
    P_\mathrm{fp}(\phi_e, \Pi, F, \mathcal{N}) = \delta(\mathcal{N}),
\end{equation}
where the subscript $\mathrm{fp}$ refers to {first passage},}
  $\mathcal{N}$ refers the stochastic time measured at the end of inflation and  {$\tilde{\Gamma}(N) \delta(N - N') \equiv \langle \xi_{{\phi}}(N) \xi_{{\Pi}}(N') \rangle$.} 
Here, all variables are considered as initial values, i.e. $\phi=\phi_\mathrm{ini}, \Pi=\Pi_\mathrm{ini}$ and $F=F_\mathrm{ini}$, and we identify $F = N - N_{\text{cl}}$ to shift the origin of time. Hereafter, we remove the subscript $ini$ for simplicity.


Solving the adjoint Fokker-Planck equation directly can be computationally challenging. A more efficient approach involves transforming the equation into its characteristic form, which is equivalent to applying a Fourier transformation to the adjoint Fokker-Planck equation.
This method significantly simplifies the handling of boundary conditions.
As demonstrated in \cite{Pattison:2021oen}, this formalism offers notable advantages, particularly in computing non-Gaussian statistics of cosmological perturbations.

Accordingly, performing the Fourier transform of Eq.~\eqref{adjoint} leads to
the  following characteristic equation,
\begin{equation}
\begin{split} \label{char}
    -i t \chi = &
    - \left( 3\Pi + \frac{3V'(\phi)}{V(\phi)} \right) \frac{\partial \chi}{\partial \Pi}
    + \Pi \frac{\partial \chi}{\partial \phi}
    + \frac{\partial \chi}{\partial F} \\
    & + \frac{f(F)^2}{2} \frac{\partial^2 \chi}{\partial \phi^2}
    + \frac{g(F)^2}{2} \frac{\partial^2 \chi}{\partial \Pi^2}
    + \Gamma(F) \frac{\partial^2 \chi}{\partial \phi \partial \Pi},
\end{split}
\end{equation}
where
 {
\[
\chi(t) \equiv  \frac{1}{\sqrt{2\pi}} \int_{-\infty}^{\infty} P_\mathrm{fp}(\phi_e, \Pi, F, \mathcal{N}) e^{i t \mathcal{N}} d\mathcal{N}, \quad  \quad \chi \big|_{\phi = \phi_e} = 1,
\]
}
so  $\chi(t)$ represents the Fourier transform of PDF. To simplify the calculations, we define and replace any function $\tilde{A}$,
 which appears in the  Eq.~\eqref{adjoint}, with
 \begin{equation}\label{Atilde}
  A(F) \equiv \tilde{A}(F + N_{\text{cl}}).
 \end{equation}

To solve Eq.~\eqref{char}, we use the method of characteristics as outlined in \cite{Pattison:2021oen}. In this approach one needs to introduce some new variables, here including $u,~c,~\text{and} ~\tilde{c}$ in which  $u$ satisfies the following chain rules,
\begin{equation}\label{characu}
    \frac{d}{du} \equiv  \frac{d\Pi}{du} \frac{\partial}{\partial \Pi}
    + \frac{d\phi}{du} \frac{\partial}{\partial \phi}
    + \frac{dF}{du} \frac{\partial}{\partial F},
\end{equation}
while $c,~\text{and} ~\tilde{c}$ will be determined in the following.

Applying this method,  Eq.~\eqref{char} is transformed into,
\begin{equation} \label{charac}
  -  i t \chi = \frac{d\chi}{du}
    + \frac{f(F)^2}{2} \frac{\partial^2 \chi}{\partial \phi^2}
    + \frac{g(F)^2}{2} \frac{\partial^2 \chi}{\partial \Pi^2}
    + \Gamma(F) \frac{\partial^2 \chi}{\partial \phi \partial \Pi} .
\end{equation}
Combining Eqs.~\eqref{char}, \eqref{charac} and ~\eqref{characu}, leads to the  following characteristic equations,
\begin{equation} \label{phase}
\begin{split}
    \frac{d\Pi}{du} &= - \left( 3\Pi + \frac{3V'(\phi)}{V(\phi)} \right), \\
    \frac{d\phi}{du} &= \Pi, \\
    \frac{dF}{du} &= 1.
\end{split}
\end{equation}
These equations have the following closed-form solutions,
\begin{equation} \label{sol}
\begin{split}
    \phi &= \phi(u, {c_1}), \\
    \Pi &= \Pi(u, {c_2}), \\
    F &= F(u, \tilde{c}) = u + \tilde{c},
\end{split}
\end{equation}
where $c_1$, $c_2$ and $\tilde{c}$ are integration constants determined by {initial/boundary} conditions.

Since we have three equations with four variables we need to fix one of them. To this end we fix $c_2$ for instance using the 
condition $N_{\text{cl}}(\phi_{e}, \Pi)=0$, 
wherein  $\phi_{e}$ refers the value of the field at the end of inflation.
One notices that $\tilde{c}$ is fixed by setting $F = -N_{\text{cl}}$.
Therefore, Eq. \eqref{sol} can be rewritten as
\begin{equation} \label{sol1}
\begin{split}
    \phi &= \phi(u, {c}), \\
    \Pi &= \Pi(u, {c}), \\
    F &= F(u, \tilde{c}) = u + \tilde{c}.
\end{split}
\end{equation}

\vspace{0.5cm}

\subsection{Moments of \texorpdfstring{$\mathcal{N}$}{N}}

In the stochastic $\delta {N}$ formalism, we are primarily interested in the moments of $\mathcal{N}$ which are the building blocks to calculate the power spectrum and bispectrum \cite{Fujita:2014tja, Vennin:2015hra, Firouzjahi:2018vet}.  Specifically, the power spectrum of curvature perturbation is given by,
\begin{equation}\label{PS-SDN}
\mathcal{P}_\calR=\frac{d~ \delta \mathcal{N}^2}{d~\langle\mathcal{N}\rangle}\,,
\end{equation}
where the variance is defined via $\delta \mathcal{N}^2 \equiv \langle \mathcal{N}^2 \rangle - \langle \mathcal{N} \rangle^2$. Furthermore, 
note that  the moments of $\mathcal{N}$ are extracted from $\chi$ as follows, 
\begin{equation} 
\label{moments}
    \left. \frac{d^n \chi}{dt^n} \right|_{t = 0} = i^n \langle \mathcal{N}^n \rangle.
\end{equation}

Below, we outline the procedure to calculate $\langle \mathcal{N} \rangle$
and  $\delta \mathcal{N}^2 $. Using Eqs.~\eqref{charac} and (\ref{moments}), the first and second moments satisfy the following equations,
\begin{equation} \label{n}
    \frac{d \langle \mathcal{N} \rangle}{du}
    + \frac{f(F)^2}{2} \frac{\partial^2 \langle \mathcal{N} \rangle}{\partial \phi^2}
    + \frac{g(F)^2}{2} \frac{\partial^2 \langle \mathcal{N} \rangle}{\partial \Pi^2}
    + \Gamma(F) \frac{\partial^2 \langle \mathcal{N} \rangle}{\partial \phi \partial \Pi}
    = -1,
\end{equation}
\begin{equation} \label{n2}
    \frac{d \langle \mathcal{N}^2 \rangle}{du}
    + \frac{f(F)^2}{2} \frac{\partial^2 \langle \mathcal{N}^2 \rangle}{\partial \phi^2}
    + \frac{g(F)^2}{2} \frac{\partial^2 \langle \mathcal{N}^2 \rangle}{\partial \Pi^2}
    + \Gamma(F) \frac{\partial^2 \langle \mathcal{N}^2 \rangle}{\partial \phi \partial \Pi}
    = -2 \langle \mathcal{N} \rangle.
\end{equation}

We impose the boundary condition $\langle \mathcal{N}^i \rangle \big|_{\phi = \phi_e} = 0$ where $\phi_e$ is the point of end of inflation which is fixed  in the field space.

Correspondingly, from the combination of the above two equations for 
$\langle \mathcal{N} \rangle$ and $ \langle \mathcal{N}^2 \rangle$, 
the equation for the evolution of the variance $\delta \mathcal{N}^2 = \langle \mathcal{N}^2 \rangle - \langle \mathcal{N} \rangle^2$ is given by, 
\begin{equation} \label{dn2}
\begin{split}
    \frac{d \delta \mathcal{N}^2}{du} = &
    - \frac{f(F)^2}{2} \left( \frac{\partial^2 \langle \mathcal{N}^2 \rangle}{\partial \phi^2}
    - 2 \langle \mathcal{N} \rangle \frac{\partial^2 \langle \mathcal{N} \rangle}{\partial \phi^2} \right)
    - \frac{g(F)^2}{2} \left( \frac{\partial^2 \langle \mathcal{N}^2 \rangle}{\partial \Pi^2}
    - 2 \langle \mathcal{N} \rangle \frac{\partial^2 \langle \mathcal{N} \rangle}{\partial \Pi^2} \right) \\
    & - \Gamma(F) \left( \frac{\partial^2 \langle \mathcal{N}^2 \rangle}{\partial \phi \partial \Pi}
    - 2 \langle \mathcal{N} \rangle \frac{\partial^2 \langle \mathcal{N} \rangle}{\partial \phi \partial \Pi} \right).
\end{split}
\end{equation}
In the drift-dominated limit, Eqs.~\eqref{n}, \eqref{n2}, and~\eqref{dn2} can be solved iteratively.  We proceed to derive both the leading-order (LO) and next-to-leading-order (NLO) contributions to the evolution of \( \langle \mathcal{N} \rangle \) and \( \langle \mathcal{N}^2 \rangle \), highlighting their consistency with the classical and quantum expectations.

\subsection{Leading Order Analysis}

At leading order (LO), we neglect stochastic fluctuations by omitting the diffusion terms. The resulting deterministic evolution equations for the first and second moments of \( \mathcal{N} \) are,
\begin{equation}
\begin{aligned}
  \frac{d\langle \mathcal{N} \rangle_{\text{LO}}}{du} &= -1, \\
  \frac{d\langle \mathcal{N}^2 \rangle_{\text{LO}}}{du} &= -2 \langle \mathcal{N} \rangle_{\text{LO}}.
\end{aligned}
\end{equation}

Solving these coupled equations yields,
\begin{equation} \label{LOsol}
\begin{aligned}
  \langle \mathcal{N} \rangle_{\text{LO}} &= -u + d_1, \\
  \langle \mathcal{N}^2 \rangle_{\text{LO}} &= (-u + d_1)^2 + d_2,
\end{aligned}
\end{equation}
where \( d_1 \) and \( d_2 \) are two constants of integration  determined by {initial/boundary} conditions which depend on the phase-space coordinate
\( \phi ~\text{and}~ \Pi \) but do not depend  to the parameter $u$. Thus,
\begin{equation} \label{c1c2}
\begin{aligned}
  d_1 &= d_1\big(\phi, \Pi\big), \\
  d_2 &= d_2\big(\phi, \Pi\big),
\end{aligned}
\end{equation}
where $d_1$ and $d_2$ are fixed on the  initial flat surfaces.

In the classical limit, and in the absence of the  quantum fluctuations, Eq.~\eqref{LOsol} corresponds to the deterministic dynamics of the scalar field in an inflationary background. Specifically, we identify
\begin{equation}
  \langle \mathcal{N} \rangle_{\text{LO}} = N_{\text{cl}},
  \quad \quad
  \langle \mathcal{N}^2 \rangle_{\text{LO}} = N_{\text{cl}}^2,
\end{equation}
where $N_{\text{cl}}$  is the classical number of e-folds until the point of end of inflation fixed by the boundary value $\phi=\phi_e$, solved by  the background Klein-Gordon equation.


\subsection{NLO Corrections in Variance and Power Spectrum}
\label{NLO0}

Now we examine the next-to-leading order (NLO) corrections within the stochastic \(\delta N\) formalism. These corrections allow for the computation of quantum back-reaction effects encoded in the stochastic noises, capturing deviations from classical dynamics.
Our aim is to calculate the power spectrum in the presence of interacting noises.


In the presence of stochastic fluctuations, the equation for the variance \(\delta\mathcal{N}^2\) at NLO order receives corrections from the quantum noise terms.  First, note that in the iteration approach, the terms in the bracket in  
Eq. \eqref{dn2} are calculated at the LO order, yielding
 \[{\left( {\frac{{{\partial ^2}\langle {{\cal N}^2}\rangle }}{{\partial {\phi ^2}}} - 2\langle {\cal N}\rangle \frac{{{\partial ^2}\langle {\cal N}\rangle }}{{\partial {\phi ^2}}}} \right)_{\text{LO}}} = \left( {\frac{{{\partial ^2}N_{cl}^2}}{{\partial {\phi ^2}}} - 2{N_{cl}}\frac{{{\partial ^2}{N_{cl}}}}{{\partial {\phi ^2}}}} \right) =2 {\left( {\frac{{\partial {N_{{\rm{cl}}}}}}{{\partial \phi }}} \right)^2}.\] Then by substituting this expression into Eq. \eqref{dn2}, the  equation for the NLO corrections is given by, 
\begin{equation}
\label{NLOvariance}
\frac{d\delta\mathcal{N}^2}{du} = -\left[
f(F)^2 \left(\frac{\partial N_{\text{cl}}}{\partial\phi}\right)^2 +
g(F)^2 \left(\frac{\partial N_{\text{cl}}}{\partial\Pi}\right)^2 +
2\Gamma(F) \left(\frac{\partial N_{\text{cl}}}{\partial\phi} \frac{\partial N_{\text{cl}}}{\partial\Pi}\right)
\right],
\end{equation}
where  \(f(F)\), \(g(F)\), and \(\Gamma(F)\) encode the field-dependent noise amplitudes in the phase space, while \(F\) representing a generic auxiliary field, which here can be considered as initial condition \(F=-N_{\text{cl}}\). It should be noted that in the single field model of inflation with a Gaussian noise,  $f(F)= H/2\pi$ which is obtained by solving  the free Mukhanov-Sasaki equation at the initial flat hypersurface.  {Furthermore, the  contributions  \(g(F)\) and \(\Gamma(F)\) are higher orders in terms of $\epsilon_H$  and they play no role in our analysis.}

{We would like to solve Eq. \eqref{NLOvariance} to obtain the power spectrum
at the end of inflation  $N=0$.} For this purpose, we set  \(F = -N_{\text{cl}}\) which leads to the following equation (after discarding the
sub-leading contributions  \(g(F)\) and \(\Gamma(F)\)),
\begin{equation}
\label{stochastic_power}
\frac{d\delta\mathcal{N}^2}{du} = -
\tilde{f}(0)^2 \left(\frac{\partial N_{\text{cl}}}{\partial\phi}\right)^2,
\end{equation}
where \(\tilde{f}(0)\) is the amplitude of  the noise function $\xi_\phi$ at
$N=0$ where the correlators are measured.

We consider the case where the  background trajectory is constrained via  \(d (d_1 )= 0\) with $d_1$ given in Eq.~\eqref{LOsol}.  It can be seen that $d_1$ is a function of the parameter $c$. Furthermore, it can be demonstrated that $\Pi_{\text{end}}$ also depends on $c$. Consequently, the condition $d(d_1) = 0$ is equivalent to $d\Pi_{\text{end}} = 0$, which corresponds to choosing a homogeneous energy density at the end of inflation when $\phi=\phi_e$.
Thence the mean e-folding number simplifies to,
\begin{equation}
d\left<\mathcal{N}\right>_\text{LO} = -du \, .
\end{equation}
Combining this with Eq.~\eqref{stochastic_power}, we finally obtain,
\begin{equation}
\label{dnvariance}
\frac{d\delta\mathcal{N}^2}{d\left<\mathcal{N}\right>_\text{LO}} = -\frac{d\delta\mathcal{N}^2}{du} =
\tilde{f}(0)^2 \left(\frac{\partial N_{\text{cl}}}{\partial\phi}\right)^2 \, .
\end{equation}

Eq. \eqref{dnvariance}  shows that the stochastic $\delta N $ formalism accurately reproduces the power spectrum derived from classical trajectory $N_{\text{cl} }(\phi)$.  However, the important point is that the amplitude of the noise  $\tilde{f}(0)^2$ should be calculated independently 
for the interacting noise.  As mentioned previously, note that for the slow-roll setup, we have $\tilde{f}(0)= H/2\pi$.

The analysis yielding to Eq. (\ref{dnvariance}) was general. As a concrete  example, in the next section we calculate $\tilde{f}(0)^2$ for the
specific setup SR-USR-SR which is employed for PBHs formation in literature. 

\section{ Quantum Loop Corrections in Noises}\label{sec: Sec4}

From the previous analysis yielding to  Eq. (\ref{dnvariance}), we have concluded that the stochastic $\delta N$ formalism can be used to calculate the power spectrum even when the noises are not free and involve interactions. The effects
of interaction are captured in the amplitude of noise $\tilde{f}(0)$. However, to calculate this amplitude, we have to use the first principle QFT analysis. Note that a similar analysis are performed even in simple slow-roll scenarios where one has to solve the Mukhanov-Sasaki equation for the free theory which yields to the standard result $\tilde{f}(0)= (H/2\pi)$ on the initial flat hypersurface. However, in the presence of interactions, the analysis are more complicated and one has to employ the perturbative in-in formalism to calculate the amplitude of the interacting noise.


The correlators of stochastic noise  can be computed from the two-point function of the field fluctuations. Specifically, using Eq. (\ref{eq:sigma}) as the starting definition of the noise, the two-point function  of the noise is given by,
\begin{equation}
\label{correlation}
\big\langle\xi_\phi(N)\xi_\phi(N')\big\rangle = \frac{1}{6\pi^2} \frac{dk_\varepsilon^3}{dN}
\big\langle\delta\phi_{k_{\varepsilon_1}} \delta\phi_{k_{\varepsilon_2}}\big\rangle \delta(N - N'),
\end{equation}
where \(k_\varepsilon \equiv \varepsilon a H\) defines a coarse-graining scale.
Also, note that the expectation $\langle\cdots \rangle$ is with respect to the vacuum of the full theory containing the interactions 
which is different than the free Bunch-Davies vacuum
$|0\rangle$.

Our job is now to calculate the two-point function  \(\left<\delta\phi_{k_1} \delta\phi_{k_2}\right>\). This is where the in-in formalism comes to the rescue.
{More specifically \cite{Weinberg:2005vy},
\begin{equation}
\label{in-in-int}
\left\langle\delta\phi_{k_1} \delta\phi_{k_2}\right\rangle =
\left\langle 0\Big|
\overline{\mathrm{T}} \exp\left(i \int_{-\infty}^{\tau_0} d\tau' H_{\text{int}}(\tau') \right)
\delta\phi_{k_1} \delta\phi_{k_2}
\mathrm{T} \exp\left(-i \int_{-\infty}^{\tau_0} d\tau' H_{\text{int}}(\tau') \right)
\Big|0 \right\rangle,
\end{equation}}
where \(H_{\text{int}}\) is the interaction Hamiltonian for field perturbations while
$T$ and $\bar{T}$ denote the time and anti-time ordering operators respectively. 

To perform the in-in analysis we need the interaction Hamiltonian.
Usually, calculating the interaction Hamiltonian in perturbation theory is a difficult task. The cubic order Hamiltonian is calculated by Maldacena  \cite{Maldacena:2002vr} while the full quartic and higher orders Hamiltonian are not known. A useful approach to bypass this difficulty is to employ the effective field theory (EFT) of inflation  \cite{Cheung:2007st, Cheung:2007sv}. The EFT formalism is useful in the decoupling limit when one neglects the gravitational back-reactions and the perturbations in the lapse and shift functions are discarded. As discussed in Section \ref{sec:stochastic_inflation} this is also the limit where we employ the stochastic $\delta N$ formalism. 
The EFT formalism is employed to calculate the loop correction in power spectrum in \cite{Firouzjahi:2023aum}. We do not repeat the detail analysis here and refer the reader for further details to \cite{Firouzjahi:2023aum}.

The interaction Hamiltonians in EFT formalism are expressed in terms of the Goldstone boson $\pi$ which captures the breaking of the four-dimensional time reparametrization invariance. On the other hand,  we are interested in the comoving curvature perturbation $\calR$.
At linear order,  $\calR$ is related to $\pi$ via $\calR = - H \pi$ while there are non-linear corrections at higher orders. However, since we measure the power spectrum at the end of attractor phase, during the final SR phase, all these non-linear terms are irrelevant and the leading linear relation is enough. On the other hand, during the SR attractor phase $\calR$ is related to $\delta \phi$ via the usual relation
$\calR = -H\frac{ \delta \phi}{\dot \phi}$. Combining these two expressions we simplify find the following linear relation in the attractor phase,
\ba
\label{phi-pi}
\delta \phi = \dot \phi \pi \, .
\ea
Correspondingly, using Eq. (\ref{correlation}),
the two-point function of $\xi_\phi$ is related to the two-point function of $\pi$ and $\calR$ as follows,
{\begin{equation}
\label{phiphi}
\begin{split}
\left< \xi_\phi(N) \xi_\phi(N') \right>  &= \frac{\dot \phi^2}{6\pi^2} \frac{dk_{\varepsilon_1}^3}{dN}
\big<(\pi_{k_{\varepsilon_1}}  )(\pi _{k_{\varepsilon_2}} )\big> \delta(N - N')\\
& = \frac{1}{{6{\pi ^2}}}\frac{{dk_{\varepsilon_1}^3}}{{dN}}2\epsilon_H (\tau )\big\langle {{\mathcal{R}_{{k_{\varepsilon_1}}}}{\mathcal{R}_{{k_{\varepsilon_2}}}}} \big\rangle \delta (N - N') \\
&\equiv 2 \epsilon_H(N) \left( \mathcal{P}^{\small{(0)}}_{\calR}+ \Delta \mathcal{P_\calR} \right) \delta(N - N'),
\end{split}
\end{equation}}
in which the dimensionless  power spectrum $\mathcal{P}_{\calR}$ is defined via,
\ba
\label{Power-calP}
\mathcal{P}_{\calR} \equiv \frac{k^3}{2 \pi^2} | \calR_k|^2 = \frac{H^2}{8 \pi^2  \epsilon_H M_P^2}.
\ea
In our notation, $\mathcal{P}^{\small{(0)}}_{\calR}$ represents the power spectrum in the free theory while \(\Delta \mathcal{P_\calR}\) encapsulates the quantum one-loop  corrections to the power spectrum.

Combining Eq.  \eqref{f-ftilde} with \eqref{phiphi},  the amplitude of the noise
is formally given by,
\begin{equation}
\label{sqf}
    \tilde{f}(N)=\sqrt{2 \epsilon_H(N) \left( \mathcal{P}^{\small{(0)}}_{\calR} + \Delta \mathcal{P_\calR} \right)}.
\end{equation}
 In particular, for a free theory with no loop correction, \(\Delta \mathcal{P_\calR}\)=0 and the above expression yields,
\ba
\tilde f(N) = \frac{H}{2 \pi} \, , \quad \quad  \big( \Delta \mathcal{P_\calR}=0\big) , 
\ea
as expected.

Now, having formally calculated the loop corrections into the amplitude of the stochastic noise,  the curvature perturbation power spectrum predicted by stochastic $\delta N$ formalism from Eq. (\ref{dnvariance}) is given by,
\begin{equation}
\label{Pzeta}
\mathcal{P}_\calR = 2 \epsilon_H(0) \left( \frac{\partial N_c}{\partial \phi} \right)^2 \left( \mathcal{P}^{\small{(0)}}_{\calR}  + \Delta \mathcal{P_\calR} \right).
\end{equation}
Of course, in order for the stochastic $\delta N$ result to be
consistent with the full in-in analysis we require that
$2 \epsilon_H(0) \left( \frac{\partial N_c}{\partial \phi} \right)^2=1$ which is indeed
the case based on the definition of $\epsilon_H$. 

Finally, using the definition of power spectrum (\ref{Power-calP}),  the amplitude of the noise in Eq. (\ref{sqf}) is simplified to,
\begin{equation}
\label{noise-0}
\tilde{f}(N)=\frac{H}{2 \pi} \sqrt{1 + \frac{\Delta \mathcal{P_\calR}}{\mathcal{P}^{\small{(0)}}_{\calR}} }.
\end{equation}
This is our main result in this section, capturing the loop corrections in the amplitude of the noise, albeit formally.

The above discussion was general. Now to be specific, we consider the  SR-USR-SR model which is employed for PBHs formation. Our goal is to calculate the effects of the interaction in the amplitude of the noises $\tilde{f}(0)$ associated to the  long CMB modes.

\begin{figure}[t]
	\centering
	\includegraphics[ width=.6\linewidth]{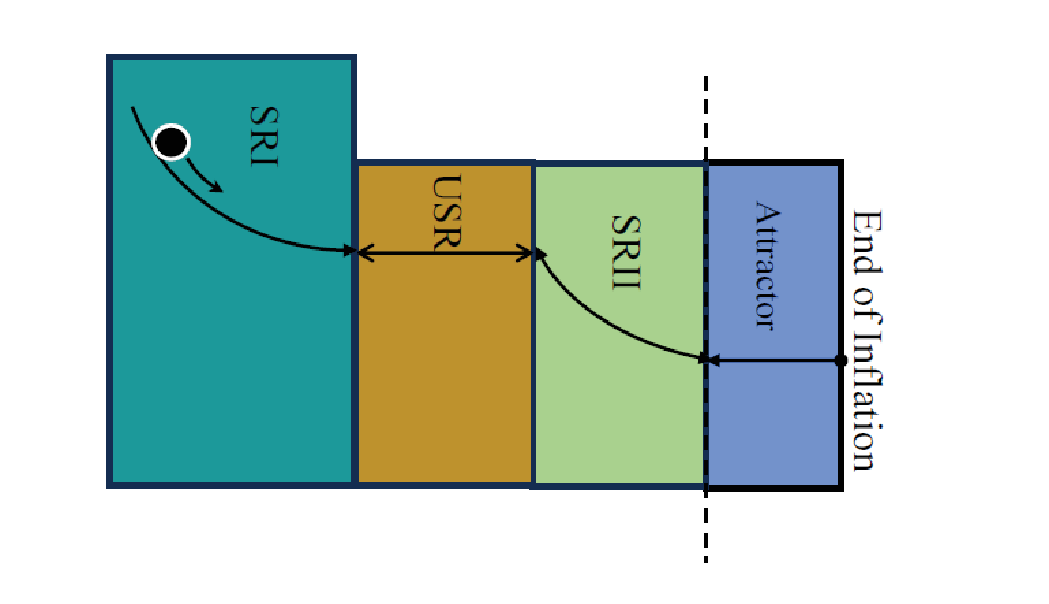}
	\caption{ A schematic view of SR-USR-SR inflationary model. The corrections to the amplitude of noise are calculated  during the attractor phase using the QFT in-in analysis.  The stochastic approach is employed afterwards to study the correlators  of the CMB modes.
}
\label{fig1}
\end{figure}

\subsection{Example: SR-USR-SR setup}\label{SUS1}

Here we briefly review the  SR-USR-SR setup, for a schematic view see Fig. \ref{fig1} and  for further details see \cite{Firouzjahi:2023aum}. The potential $V(\phi)$ is such that the first stage is a SR phase during which the long CMB modes leave the horizon. The second stage is in the form of USR where the potential is flat and the first SR parameter falls off exponentially $\epsilon_{H}
\propto a(\tau)^{-6}$. As a result, the curvature perturbation $\calR$ grows
rapidly on superhorizon scales, $\calR \propto a(\tau)^3$.
The USR phase is limited during the interval $\tau_i < \tau < \tau_e$ during which  the second SR parameter, $\eta,$ is nearly constant with $\eta\simeq -6$.
The system enters the final attractor SR phase after $\tau > \tau_e$ and inflation ends followed by reheating. 

The cosmological correlators are measured at the time of end of inflation. However, since the system reaches the attractor phase 
in the final stage, the curvature perturbation is frozen on superhorizon scales and the correlators can be measured at anytime during the final attractor phase, see Fig.  \ref{fig1}.  Correspondingly, one can calculate the amplitude of the noise anytime during the attractor phase employing the in-in formalism. Once the amplitude of the noise is obtained, then one employs the stochastic $\delta N$ formalism  to calculate the power spectrum and bispectrum for the long CMB modes.

During  the final SR phase the inflaton equation of motion is approximated via,
\begin{equation}
\label{EoM00}
\frac{d^2 \phi}{dN^2} + 3 \frac{d\phi}{dN} + 3 M_P \sqrt{2\epsilon_V} \simeq 0 \, ,
\qquad \text{with} \qquad 3 M_P^2 H^2 \simeq V(\phi_e),
\end{equation}
where the slow-roll parameter \(\epsilon_V\) is defined as
\(\epsilon_V \equiv \frac{M_P^2}{2} \left(\frac{V'}{V}\right)^2\), evaluated at the second USR phase when the system has reached to its attractor phase.
The solution of the above equation is given by  \cite{Cai:2018dkf, Firouzjahi:2023aum},
\begin{equation}
    \phi(N) = C_1 - \sqrt{2\epsilon_V}N + C_2 e^{-3N},
\end{equation}
where the coefficients $C_1$ and $C_2$ are determined by the initial 
conditions,
\begin{equation}
\begin{split}
    C_1 &= \frac{\Pi}{3} + \frac{\sqrt{2\epsilon_V}}{3} + \phi, \\
    C_2 &= -\frac{\Pi}{3} - \frac{\sqrt{2\epsilon_V}}{3}.
\end{split}
\end{equation}
Neglecting the decaying contribution, from the above solution we find that,
\begin{equation}
\label{Ncd}
\frac{\partial N_c}{\partial \phi} \simeq \frac{1}{\sqrt{2 \epsilon_V}} + \mathcal{O}\left(e^{-3N_c}\right).
\end{equation}
Following the discussion below Eq. (\ref{Pzeta}) and noting that to leading order 
in slow-roll expansion, $\epsilon_V \simeq \epsilon_H$,  this shows that indeed the  power spectrum from stochastic $\delta N$ formalism  aligns with the result obtained from the one-loop in-in formalism as expected.

A key effect is the transition from the USR stage to the final attractor phase. If this transition is mild, then the mode function keeps evolving during the subsequent SR phase while for a sharp transition, the mode function is frozen quickly after the USR phase. This effect is parameterized by the sharpness parameter $h$ as follows \cite{Cai:2018dkf},
\begin{equation}
\label{h-def}
h \equiv  -6 \sqrt{\frac{\epsilon_V}{\epsilon_e}} \, ,
\end{equation}
in which $\epsilon_e$ is the value of $\epsilon_H $ at $\tau=\tau_e$ while $\epsilon_V$ is the value of $\epsilon_H$ at the final stage of inflation when the system has reached the attractor phase.

In this picture,  $\epsilon_H$ is  smooth  across the transition  but  $\eta$ has a jump  at $\tau=\tau_e$. During the USR phase and prior to the transition at $\tau=\tau_e$,  $\eta=-6$ while immediately after the transition  $\eta= -6-h$.
Therefore,  near the transition point one can use the following approximation
\cite{Cai:2018dkf}, 
\ba
\eta = -6 - h \theta(\tau -\tau_e) \quad \quad  \tau_e^- < \tau < \tau_e^+ \, ,
\ea
leading to,
\ba
\label{eta-jump}
\frac{d \eta}{d \tau} = - h \delta (\tau -\tau_e)  \, ,  \quad \quad  \tau_e^- < \tau < \tau_e^+ \, .
\ea
We work in the limit of  sharp transition where $|h| \gg 1$ in which the mode function quickly freezes to its attractor value after the transition. A particular case of sharp transition is when $h=-6$, as considered for example in \cite{Kristiano:2022maq},  during which $\epsilon_H$ in the final SR phase is fixed to $\epsilon_e$ .

The cubic and quartic interaction Hamiltonians relevant for one-loop correction of
power spectrum is given by \cite{Firouzjahi:2023aum},
\begin{align}
\label{H3}
\mathbf{H}_3 &= - M_P^2 H^3 \eta \epsilon_H a^2 \int d^3x \left( \pi \pi'^2 + \frac{1}{2} \pi^2 \nabla^2 \pi \right), \\
\label{H4}
\mathbf{H}_4 &= \frac{M_P^2}{2}\epsilon_H \int d^3x \biggl[
\left(H^4 \eta^2  a^2 - \eta' H^3 a\right) \pi^2 \pi'^2 
+ \left(H^4 \eta^2  a^2 + \eta' H^3  a\right) \pi^2 (\nabla \pi)^2 \biggr],
\end{align}
where a prime here denotes the derivative with respect to the conformal time.  An important effect is that the parameter $\eta$ has a jump at the point $\tau=\tau_e$ where the USR phase is attached to the second SR phase.
As parameterized by Eq. (\ref{eta-jump}), this induces a localized term of the form $\delta (\tau-\tau_e)$ which should be taken into account when calculating the loop corrections.

In calculating the loop correction using in-in master formula Eq. (\ref{in-in-int}), there are two different types of Feynman diagrams at one-loop order which should be taken into account. The first diagram involves a one-vertex quartic Hamiltonian which is easier to handle analytically. The second diagram involves  two vertices of the cubic Hamiltonians. The analysis associated to this latter diagram is more complicated as it involves two  nested times in-in integrals. 
Performing the in-in analysis \cite{Firouzjahi:2023aum}, the fractional one-loop correction on long CMB modes induced from the small scale modes which become superhorizon during the USR phase is given by, 
\begin{equation}
\label{total-power-fractional}
\frac{\Delta \mathcal{P}_\calR}{\mathcal{P}^{\small{(0)}}} = \frac{6}{h} ( h^2+ 24 h + 180) \Delta N e^{6 \Delta N} \mathcal{P}^{\small{(0)}}_{\calR}  \, ,
\end{equation}
 here $\mathcal{P}^{\small{(0)}}_{\calR}$ represents the power spectrum on CMB scales which is fixed by COBE normalization to be $\mathcal{P}^{\small{(0)}}_{\calR}\simeq 2.1 \times 10^{-9}$. In addition, $\Delta N$ is the duration of the USR phase which is usually around 2-3 e-folds to generate the PBHs of desired mass scales.  If $\Delta N$ is too large and the transition is sharp, then $\Delta \mathcal{P}_\calR \sim \mathcal{P}^{\small{(0)}}_{\calR}$ and the perturbativity of the system under the loop correction is lost as advocated originally in  \cite{Kristiano:2023scm, Kristiano:2022maq}. This question has attracted significant interests in literature, see \cite{ Kristiano:2023scm,Firouzjahi:2024sce,Riotto:2023hoz, Riotto:2023gpm, Choudhury:2023vuj,  Choudhury:2023jlt,  Choudhury:2023rks, Choudhury:2023hvf,Choudhury:2024one, Choudhury:2024aji, Firouzjahi:2023aum, Motohashi:2023syh, Firouzjahi:2023ahg, Tasinato:2023ukp, Franciolini:2023agm, Firouzjahi:2023btw, Maity:2023qzw, Cheng:2023ikq, Fumagalli:2023loc, Nassiri-Rad:2023asg, Meng:2022ixx, Cheng:2021lif, Fumagalli:2023hpa,  Tada:2023rgp,  Firouzjahi:2023bkt, Iacconi:2023slv, Davies:2023hhn, Iacconi:2023ggt,  Kristiano:2024vst, Ballesteros:2024zdp, Kawaguchi:2024lsw, Braglia:2024zsl, Braglia:2025qrb,Braglia:2025cee, Firouzjahi:2024psd, Caravano:2024moy, Caravano:2024tlp, Caravano:2025diq, Saburov:2024und,  Sheikhahmadi:2024peu, Inomata:2024lud, Kawaguchi:2024rsv, Fumagalli:2024jzz, Kristiano:2025ajj, Inomata:2025pqa, Inomata:2025bqw}.
 
Plugging Eq. (\ref{total-power-fractional}) into Eq. (\ref{noise-0}), the amplitude of the noise with the effects of the loop correction included is given by, 
\begin{equation}
\label{noise-USR}
\tilde{f}(0)=\frac{H}{2 \pi} \sqrt{1 +  \frac{6}{h} ( h^2+ 24 h + 180) \Delta N e^{6 \Delta N} \mathcal{P}^{\small{(0)}}_{\calR} }.
\end{equation}
The above expression shows that the correction in the amplitude of the noise exponentially depends on the duration of the USR phase $\Delta N$. Furthermore, it grows approximately linearly with the sharpness parameter $h$.
 As argued in \cite{Firouzjahi:2023aum}, for $|h| \gg 1$, 
 the sharper is the transition to the final SR phase, the larger is the loop corrections and correspondingly the more significant the corrections in the amplitude of the noise. 
 
\begin{figure}[t]
	\centering
	\includegraphics[ width=.6\linewidth]{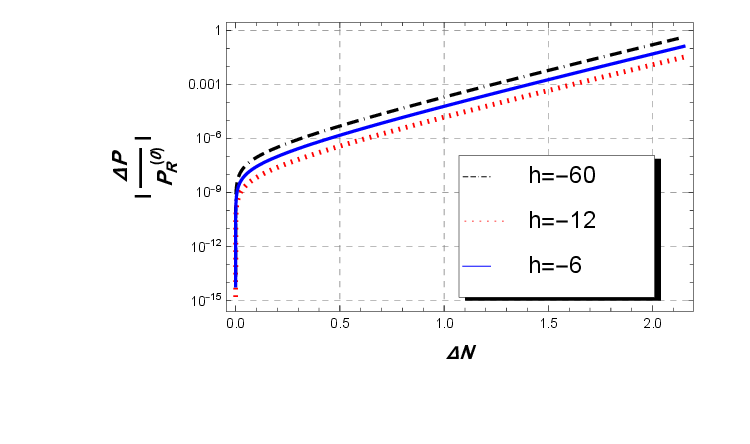}
	\vspace{-0.5 cm}
	\caption{ A schematic diagram of  the fractional one-loop correction as a function of the duration of the USR phase $\Delta N$ for different values of sharpness parameter $h$ according to Eq. \eqref{total-power-fractional}. The exponential dependence on $\Delta N$ is evident. Due to non-linear dependence on $h$, the curve for $h=-6$ is actually above the curve for $h=-12$. But for $|h|\gg 1$, the loop correction scales linearly with $h$. 
}
\label{fig2}
\end{figure}

 In Fig. \ref{fig2} we have plotted the fractional one-loop correction $|\frac{\Delta \mathcal{P}_\calR}{\mathcal{P}^{\small{(0)}}}|$ in Eq. (\ref{total-power-fractional}) as a function of the duration of the USR phase $\Delta N$ for various values of the sharpness parameter $h=-6, -12, -60$. The exponential dependence on $\Delta N$ is evident in this figure. In order for the perturbative description to be under control, we typically need $\Delta N \lesssim 2.3$. Due to non-linear dependence on $h$, the curve corresponding to $h=-6$ is actually higher than the curve corresponding to 
 $h=-12$. However, for $|h| \gg 1$, we see that the fractional loop correction increases as $h$ increases.  Note that since $h<0$, the fractional one-loop correction in Eq. (\ref{total-power-fractional}) is actually negative, $\frac{\Delta \mathcal{P}_\calR}{\mathcal{P}^{\small{(0)}}} <0$, so the amplitude of the noise is smaller than  $\frac{H}{2 \pi}$ in this SR-USR-SR setup.

\section{Implication for Fokker-Planck and Langevin Equations}

In the above analysis we have calculated the amplitude of the noise when the interaction is not negligible. The correction in noise is related to the loop corrections in power spectrum. However, translated into the stochastic $\delta N$ formalism, it is interpreted with a noise whose amplitude is modified compared to the standard value $H/2\pi$.  Correspondingly, the PDF of the adjoint Fokker-Planck equation is modified accordingly. This modification can be employed to calculate the abundance of PBHs formation.
More specifically, having calculated the correction in the amplitude of
noise, the
adjoint Fokker-Planck equation \eqref{adjoint} is rewritten as follows,
 {
\begin{equation}
\begin{split} \label{adjointm}
    \frac{\partial P_\mathrm{fp}}{\partial \mathcal{N}} = &
    - \left( 3\Pi + \frac{3V'(\phi)}{V(\phi)} \right) \frac{\partial P_\mathrm{fp}}{\partial \Pi}
    + \Pi \frac{\partial P_\mathrm{fp}}{\partial \phi}
    + \frac{\partial P_\mathrm{fp}}{\partial F} \\
    & + \epsilon_H(F + N_{\text{cl}})\Big(\mathcal{P}^{\small{(0)}}_{\calR}(F+N_{\text{cl}})+\Delta\mathcal{P}_\calR(F+N_{\text{cl}})\Big) \frac{\partial^2 P_\mathrm{fp}}{\partial \phi^2}
\end{split}
\end{equation}
}
Solving the above equation yields the modified PDF. Note that Eq. (\ref{adjointm}) is general with the correction in the amplitude of the noise captured in $\Delta\mathcal{P}_\calR$. We have specifically calculated this correction for the particular setup SR-USR-SR studied in 
Section \ref{sec: Sec4} as given in Eq. (\ref{total-power-fractional}) or equivalently in 
\eqref{noise-USR}.

Alternatively, in terms of the starting Langevin equation (\ref{Langevin}), the evolution of the field in the attractor phase, after the correction into the noise is taken into account, is written as,
\ba
\label{Langevin2}
\frac{d \phi}{d N} \simeq -\frac{V_\phi}{3 H^2} + \frac{H}{2 \pi} \sqrt{1+ \frac{\Delta \calP_\calR}{\mathcal{P}^{\small{(0)}}_{\calR}}}  \xi(N)\, .
\ea
Intuitively speaking, the above Langevin equation suggests that the
amplitude of the stochastic jump is given by $\frac{H}{2 \pi} \Big(1+ \frac{\Delta \calP_\calR}{\mathcal{P}^{\small{(0)}}_{\calR}}\Big)^{\frac{1}{2}}$.

One can use the above Langevin equation in standard stochastic process such as employing the Ito calculus to study  the first boundary crossing.  For example, as studied in \cite{Vennin:2015hra},  suppose we have a configuration where
the inflaton field is initially at the position $\phi_*$ while
inflation can end in two different endpoints, $\phi_1$ or $\phi_2$. The probability of ending the inflation by first crossing the endpoint $\phi_1$ ($\phi_2$) is denoted by $p_1(p_2)$ with the understanding that $p_1+ p_2=1$.  Following a similar analysis as in \cite{Vennin:2015hra} using the Ito lemma, the probability is calculated to be,
\ba
p_1= \frac{\int_{\phi_*}^{\phi_2}  { e^{\frac{-1}{v(x)}} }   dx   }
{ \int_{\phi_1}^{\phi_2} e^{\frac{-1}{v(x)}}   dx  } \, ,
\ea
in which now,
\ba
\label{v-phi}
v(\phi)\equiv  \big( 1 + \frac{\Delta \calP_\calR}{\mathcal{P}^{\small{(0)}}_{\calR}} \big)
 \Big(\frac{V(\phi)}{24 \pi^2 M_P^4}  \Big) \, .
\ea
We see that the correction in the noise (or diffusion term) affects the probabilities $p_i$ non-perturbatively. 

 { From this analysis  one 
may conclude  that the loop corrections computed in this work just modify  the amplitude of the noise. However, we comment that this is not a simple rescaling of the original potential, because the corrections depend on the dynamics of the sub-Hubble modes. One has to take into account the effects of the backreaction of the new effective potential so the effective description is not merely a rescaling of the potential 
but a genuine modification of the stochastic differential equation.
}

As a second example, consider the stationary PDF associated to the Langevin equation (\ref{Langevin2}) or the adjoint Fokker-Planck equation (\ref{adjointm}).
This corresponds to the case where $\partial P_{\text{stat}}/\partial N=0$, yielding to  \cite{Vennin:2015hra},
\ba
P_{\text{stat}}(\phi) \propto \frac{e^{\frac{1}{v(\phi)}}}{v(\phi)}, 
\ea
with $v(\phi)$ given in Eq. (\ref{v-phi}). 
Again, we see that a modification in the amplitude of the noise affects the PDF non-perturbatively.

As the third example, suppose that $\Delta \calP_\calR$ becomes significant, for example  for large value of $\Delta N$ in Fig.~\ref{fig2}. In this case  the diffusion term dominates over the drift term in Eq.~\eqref{adjointm}. In this diffusion-dominated regime, the probability density can be calculated to describe the statistics of PBHs formation. For example, 
for a simplified one-dimensional system (i.e. a single field model), the PDF in the diffusion-dominated regime 
is governed by the Fokker-Planck equation \cite{Pattison:2021oen,Asadi:2023flu},
\begin{equation}
\label{fokkerp}
\frac{\partial P(\phi, N)}{\partial N} = b^2 \frac{\partial^2 P(\phi, N)}{\partial \phi^2},
\end{equation}
where $P(\phi, N)$ is the probability density and $b$ the diffusion coefficient, nearly  constant, which is related to the Hubble scale and the loop correction in 
the power spectrum via, 
\begin{equation}
\label{Langevin2}
b ={ \tilde{f(0)} }= \left( \frac{H}{2\pi} \right) \sqrt{1 + \frac{\Delta \mathcal{P}_\mathcal{R}}{\mathcal{P}_\mathcal{R}^{(0)}}}\, .
\end{equation}

In the presence of two absorbing barriers at $\tilde{\phi}_\pm$, the boundary condition and initial condition are,
\begin{equation}
P(\tilde{\phi}_\pm, N) = 0, \quad \quad 
P(\phi, 0) = \delta(\phi - \tilde{\phi}),
\end{equation}
where $\tilde{\phi}$ is the initial field value. 

Solving Eq.~\eqref{fokkerp} with these conditions, the PDF for the first passage time, $P_{\text{diff}}(\mathcal{N})$, is given by  \cite{Pattison:2021oen,Asadi:2023flu}
\begin{equation}
\label{Pdiff}
P_{\text{diff}}(\mathcal{N}) = -b^2 \Big( \left. \frac{\partial P(\phi, \mathcal{N})}{\partial \phi} \right|_{\phi = \tilde{\phi}_+} - \left. \frac{\partial P(\phi, \mathcal{N})}{\partial \phi} \right|_{\phi = \tilde{\phi}_-} \Big).
\end{equation}
This yields the explicit form of the PDF in the diffusion-dominated regime
 \cite{Pattison:2021oen,Asadi:2023flu},
\begin{equation}
\label{PDFpm}
P_{\text{diff}}(\mathcal{N}) = 4 \frac{b^2 \pi}{(\tilde{\phi}_+ - \tilde{\phi}_-)^2} \sum_{m=1}^\infty m (1 - \cos(m \pi)) \sin \Big( m \pi \frac{\tilde{\phi} - \tilde{\phi}_-}{\tilde{\phi}_+ - \tilde{\phi}_-} \Big) \exp \left( - \frac{b^2 \pi^2 m^2}{(\tilde{\phi}_+ - \tilde{\phi}_-)^2} \mathcal{N} \right).
\end{equation}
Using Eq. \eqref{PDFpm}, the first and second moments of the first passage time $\mathcal{N}$ are, 
\ba
\label{phiPmin}
\langle \mathcal{N} \rangle &=& - \frac{\tilde{\phi}_+ \tilde{\phi}_-}{2 \epsilon_H} \left( \mathcal{P}_\mathcal{R}^{(0)} + \Delta \mathcal{P}_\mathcal{R} \right)^{-1}, \nonumber\\
&=& - \frac{\tilde{\phi}_+ \tilde{\phi}_-}{\big(\frac{H}{2 \pi} \big)^2}
\Big(1 + \frac{\Delta \mathcal{P}_\mathcal{R}}{\mathcal{P}_\mathcal{R}^{(0)}} \Big)^{-1}
\ea
and
\ba
\langle \mathcal{N}^2 \rangle &=& - \frac{\tilde{\phi}_+ \tilde{\phi}_-}{12 \epsilon_H^4} \left( \tilde{\phi}_-^2 - 3 \tilde{\phi}_+ \tilde{\phi}_- + \tilde{\phi}_+^2 \right) \left( \mathcal{P}_\mathcal{R}^{(0)} + \Delta \mathcal{P}_\mathcal{R} \right)^{-2}\nonumber\\
&=&  - \frac{\tilde{\phi}_+^2 \tilde{\phi}_-^2}{ \big(\frac{H}{2 \pi} \big)^4} 
\Big(1 + \frac{\Delta \mathcal{P}_\mathcal{R}}{\mathcal{P}_\mathcal{R}^{(0)}} \Big)^{-2}\Big( \frac{\tilde{\phi}_-}{3\tilde{\phi}_+}   + \frac{\tilde{\phi}_+}{3\tilde{\phi}_-} -1\Big)  \, ,
\ea
with the assumption that  the diffusion term remains approximately constant. Both $\langle \mathcal{N} \rangle$ and $\langle \mathcal{N}^2 \rangle$ can decrease or increase depending to the sign of loop corrections, indicating significant modifications to the PDF of the curvature perturbation $\calR$. Furthermore, as expected, the above results suggest that each stochastic jump has the modified amplitude  $\frac{H}{2 \pi} \Big(1+ \frac{\Delta \calP_\calR}{\mathcal{P}^{\small{(0)}}_{\calR}}\Big)^{\frac{1}{2}}$. \\

The conclusion from these examples is that  whenever the loop correction $\Delta \calP_\calR$ is significant then the deviation from the standard amplitude $H/2 \pi$ is large and one should consider the correction in the amplitude of the noise. 
Of course, a large loop correction with $|\Delta \calP_\calR/\mathcal{P}^{\small{(0)}}_{\calR}| \sim 1$ indicates that the system is non-perturbative and higher orders loop corrections become important \cite{Firouzjahi:2025gja, Firouzjahi:2025ihn}. Whether or not the perturbativity in loop correction is lost depends on the model parameters. For instance, in the example of 
SR-USR-SR setup reviewed in previous section this crucially depends on the duration of the USR phase $\Delta N$ and the  sharpness parameter $h$.

\section{Summary and Discussions}
\label{Concluding}

Stochastic $\delta N$ formalism is a powerful tool to calculate the power spectrum and bispectrum in inflationary models. One great advantage of $\delta N$ approach is that it is formulated non-linearly so this formalism can be employed for non-perturbative analysis such as calculating the tail of the PBHs formation which is a rare and non-perturbative phenomenon. However, in the standard formulation of the stochastic $\delta N$ formalism, the underlying theory is nearly free (such as  in the SR limit) so the noise on the initial flat hypersurface
is approximated by a white noise of the fixed amplitude $\frac{H}{2 \pi}$.
In this work we extended the above picture to the cases where the underlying theory is not free and, correspondingly, the initial stochastic noises have a time-dependent amplitude.

Beginning with a LO analysis in a general single field model,
we demonstrated that in the absence of stochastic noise, the expected number of $e$-folds aligns with the classical predictions of the background Klein-Gordon solution. 
Extending to NLO, we incorporated the effects of stochastic noise sourced by quantum fluctuations of the inflaton field. This {again} allowed us to derive correction to the variance of $\mathcal{N}$, which is directly tied to the power spectrum of curvature perturbations. We demonstrated that $\delta\mathcal{N}^2$ receives contributions from phase-space noise terms, with each term being proportional to derivatives of the classical number of e-folds with respect to the field variables. By evaluating the NLO corrections,  we demonstrated in Eq. \eqref{dnvariance} that the stochastic formalism reproduces the known classical $\delta N$ formula for power spectrum, validating its robustness.  However, to employ the stochastic $\delta N$ formalism in these setups, one needs the modified amplitude of the noise $\tilde f(0)$ as an input quantity. This is calculated using the QFT in-in formalism in which the loop corrections in the power spectrum is mapped to the amplitude of the noise. More specifically, we have shown that the amplitude of the noise is modified to $\frac{H}{2 \pi} \Big(1+ \frac{ \Delta {\cal P}_{\cal R} }{ {\cal P}^{{(0)}}_{{\cal R} } }\Big)^{\frac{1}{2}}$ in which  $ \frac{\Delta {\cal P}_{\cal R} }{ {\cal P}^{{(0)}}_{{\cal R} } }$ is the fractional one-loop correction in power spectrum. 

As a concrete example, we studied a three-phase model of inflation, comprising two SR phases separated by a USR regime. As the theory has non-trivial interactions in the USR phase, the amplitude of the noise is modified compared to the free theory with the  corrections in the amplitude of the noise being related to the loop correction in power spectrum. The one-loop correction in power spectrum is previously calculated using in-in formalism within EFT approach which enables us to calculate the correction in the amplitude of the noise as given in Eq. (\ref{noise-USR}). It is shown that the amplitude of the noise is exponentially sensitive to the duration of the USR phase $\Delta N$ with non-trivial dependence to  the sharpness parameter $h$ as well.   {Having said this, we comment that our stochastic formalism can not shed further light on the current debate on whether or not the small scale loop corrections can affect the large scale modes as advocated in \cite{Kristiano:2022maq}. The reason is that the stochastic formalism is an effective approach and in our study we have employed the results from the QFT in-in analysis such as in \cite{Firouzjahi:2023aum} to calculate the corrections in the initial amplitude of the noise. 
In this view, the results from our stochastic formalism are parallel with the results of \cite{Firouzjahi:2023aum} and can not be viewed as an independent insight in the current debate about the loop corrections on PBHs formation as raised in  \cite{Kristiano:2022maq}.  
}

We have rewritten the Langevin and the adjoint Fokker-Planck equations with the effects of the quantum correction in the amplitude of the noise taken into account. In the presence of large quantum corrections, the diffusion terms can become large which in turn can have important implications in PDF and the statistics of rare events such as the PBHs formation and in the studies of the first boundary crossings.  Therefore, our study  provides not only a theoretical consistency but also  a framework for computing cosmological observables in regimes where perturbative expansion in field space is subtle.

\section*{Acknowledgment}  A. N. and H. Sh.  would like to thank S. Hooshangi for useful  discussions. H. Sh. gratefully acknowledges the warm hospitality extended by North-West University (NWU) and A. Abebe during the course of this work. The work of H. F. is partially supported by INSF of Iran under the grant number  4046375. 

\appendix

\section{Pawula Theorem}
\label{AppendixB}

For a continuous {one dimensional} Markov process described by the Fokker-Planck equation  {
\begin{equation}
    \frac{\partial P(\phi,N)}{\partial N} = \sum_{k=1}^\infty \frac{(-1)^k}{k!} \frac{\partial^k}{\partial \phi^k} \left[ D^{(k)}(\phi) \, P(\phi,N) \right]\,,
\end{equation}
    }
{the Pawula's theorem \cite{Pawula1967} states that:}
\begin{enumerate}
\item {Either the sequence $D^{(1)}$, $D^{(2)}$, $D^{(3)},~\cdots$ becomes zero at the third term, or all its even terms are positive.}
\item The only consistent truncations are:
    \begin{itemize}
    \item At 1st order (pure deterministic drift).
    \item At 2nd order (standard Fokker-Planck).
    \item Infinite order (full Kramers-Moyal expansion).
    \end{itemize}
\end{enumerate}

This implies that for physical processes, the Fokker-Planck description (2nd order truncation) is either {exact} or an {approximation} whose validity must be carefully justified.


\end{document}